\documentclass[a4paper,12pt]{article}
\usepackage[utf8]{inputenc}
\usepackage{amssymb}
\def\be {\begin{equation}}
\def\ee {\end{equation}}
\def\bea {\begin{eqnarray}}
\def\eea {\end{eqnarray}}
\def\bc {\begin{center}}
\def\ec {\end{center}}
\def\bfg {\begin{figure}}
\def\efg {\end{figure}}
\def\bi {\begin{itemize}}
\def\ei {\end{itemize}}

\def\pa {\partial}

\def\a {\alpha}
\def\b {\beta}
\def\g {\gamma}
\def\m {\mu}
\def\n {\nu}
\title{Higher Derivative Terms in Three Dimensional Supersymmetric Theories }
\author{Adel Awad$^{1,3}$  and Mir Faizal$^2$ \\
$^1$Center for Theoretical Physics, British University of Egypt\\
Sherouk City 11837, P.O. Box 43, Egypt\\
$^2$Department of Physics and Astronomy, \\  University of Waterloo,   Waterloo,\\
Ontario N2L 3G1, Canada\\
$^3$Department of Physics, Faculty of Science,\\
Ain Shams University, Abbassia, Cairo 11566, Egypt\\
}
\date{}
\begin{document}

\maketitle

\begin{abstract}
In this work, we systematically analyze higher derivative terms in the supersymmetric 
effective actions for three dimensional scalar field theories using $\mathcal{N} =1$ 
superspace formalism. In these effective actions, we show that auxiliary fields do not 
propagate and their effective actions can be expressed in terms of the physical fields.  
So, the theory does not change its field content upon addition of  higher derivative terms. 
We use derivative expansion to generate four, 
five and six dimensional terms for an interacting  scalar  
field theory with $\mathcal{N} =1$ supersymmetry.
We show that along with pure fermionic and bosonic terms, there are various five and six dimensional
topological terms that mix bosonic and fermionic fields. Finally, we use these results to obtain
higher derivative topological terms in the effective action for two M2-branes.
\end{abstract}
\section{Introduction}

Three dimensional supersymmetric field theories are interesting as they
have been analyzed as examples of the $AdS_4/CFT_3$ correspondence in M-theory. The $OSp(8|4)$ symmetry of the eleven dimensional supergravity on $AdS_4 \times S_7$ is realized as $\mathcal{N} = 8$ supersymmetry of the boundary superconformal field theory.
This boundary superconformal field theory describes a system of multiple M2-branes.
Furthermore, this boundary theory is constrained
not to have any on-shell degrees of freedom coming from the gauge fields.
All these properties are satisfied by BLG theory  \cite{1, 2, 3, 4, 5}. The BLG theory only describes two M2-branes.
However, it has been possible to construct a generalization of the BLG theory called the ABJM theory \cite{abjm, ab, ab1,
ab2}. The ABJM theory is thought to describe multiple M2-branes, and it reduces to the BLG theory for two M2-branes.
Even though the ABJM theory has only $\mathcal{N} =6$ supersymmetry, it is expected that its supersymmetry might get enhanced
to full $\mathcal{N} =8$ supersymmetry  \cite{abjm2}.
Just as the higher derivative correction to the D2-brane action can be written in form of Dirac-Born-Infeld action,
it is possible to write higher derivative corrections to the ABJM theory. This can be done by writing the 
matter part of the
ABJM theory in form of a gauge covariantized   Nambu-Goto action.
It may be noted that higher derivative corrections to this non-linear extension of  the
ABJM model have also been studied \cite{hi}. It has been shown that the Mukhi-Papageorgakis higgs mechanism
  can be used to determine higher 
  derivative corrections to the BLG effective action \cite{mukhi-1}. This formalism 
  is an on-shell formalism. 

It may be noted that apart from the application to the physics of M2-branes and D2-branes, the addition of higher derivative
corrections is interesting in its own right. Recently a generic  three dimensional supersymmetric  gauge theory coupled to matter fields  has been constructed \cite{q}. Under various limits this  generic action reduces to the supersymmetric Maxwell theory,
supersymmetric Maxwell-Chern-Simons, and supersymmetric Chern-Simons theories with   matter fields. A generic three dimensional
higher derivative superfield theory for self interacting scalar superfields has also been
constructed \cite{q0}. In this analysis the self interacting higher derivative actions for real and complex scalar superfields  have been studied.

Furthermore, supersymmetric theories with higher derivative terms play an important role in various cosmological models \cite{p1}.
In the Dirac-Born-Infeld inflation, a scalar field describes the position of a brane plays the role of
the inflaton field and causes an accelerated expansion of the universe \cite{p2, p4}.
 The higher derivatives in the action cause new dynamics to arise and lead to equilateral-type non-gaussianity
 in the primordial density fluctuations \cite{p5}.
 Furthermore, higher derivatives play an important role even in ekpyrotic universes \cite{p6, p7}.
 In these universes big bang  is produced by the collision of branes moving  
 in the bulk. In this model, a  phase transition   from a   contracting  phase to an expanding phase 
 occurs for the ekpyrotic universes. This phase transition requires the violation of null energy condition.
This conditions can be violated if the sum of the pressure and the effective energy density is negative.
However, this leads to the existence of ghosts. It is possible to overcome this problem by adding 
higher derivative terms \cite{cs, cs1}. This occurs because  of ghost condensation \cite{gc, gc1}.

It is known that certain higher derivative (HD) terms in effective actions of
gauge theories with extended supersymmetries are not renormalized. The  
Seiberg-Dine terms with  $\mathcal{N} = 2$ supersymmetry 
\cite{sd,dg},  and Wess-Zumino terms in the four dimensional  super-Yang-Mills theory 
with $\mathcal{N} = 4$ supersymmetry 
\cite{tz,bs},  are examples of such terms. Furthermore, the  HD terms generated from the 
D3-brane action have been used for analysing   
  non-renormalization properties and   anomalous dimensions
of the four dimensional  super-Yang-Mills theory 
with $\mathcal{N} = 4$ supersymmetry. We hope to obtain similar results for the M2-branes, 
and this is one of the main motivations for this paper.  It may be noted that the 
HD corrections for  four dimensional field 
theories with $\mathcal{N} = 1$ supersymmetry have already  been studied \cite{s1, s2, s3, s4}. 
Such terms have also been studied using the four dimensional 
harmonic superspace \cite{h1, h2, h3, h4}. 
As M2-branes are described by a three dimensional 
superconformal field theory, we will analyse HD corrections to three dimensional 
supersymmetric field theories. So, in  this work, we analyze HD terms generated in
derivative expansion of three dimensional  supersymmetric field theories in $\mathcal{N}= 1$ superspace formalism.

In this work,  we first consider the derivative expansion of an interacting 
supersymmetric scalar  field theory.   This analysis will be performed using $\mathcal{N}=1$ 
superspace formalism. It has been argued that the addition of 
HD terms in superspace formalism can cause problems in the original theory 
\cite{g1,g2,o1,o2}. This is because  a generic higher derivative 
action  would contain terms like $\int_{superspace} (\pa \Phi)^2 \sim \int_{spacetime}(\pa F)^2$,  
(where $F$ is the auxiliary field). This will produce  kinetic terms for such auxiliary field.  Thus, 
the HD terms will add 
new unwanted degrees of freedom to the original theory. Furthermore, in certain cases, these kinetic terms
for the auxiliary field have the wrong sign, and this breaks the 
 unitarity of the original  theory. The   vacuum of the 
theory can become unstable because of the HD terms. 
However, in this paper, we are able to explicitly demonstrate that the HD terms in the 
  derivative expansion of a supersymmetric field theory, 
will only contain  non-propagating auxiliary field. So, the action for this  effective field theory can be 
always written in terms of physical fields.  
Therefore, the field content of the theory does not change upon adding 
HD terms. This is an important result as there has been a confusion regarding this point in earlier works
\cite{g1,g2,o1,o2}. 

Thus, using the superspace formalism, we obtain various four, five and six dimensional HD terms 
for an interacting scalar field theory. We show that pure 
fermionic and mixed topological terms with five and six mass dimensions
exist in the effective action of this theory. Finally, we apply these results to obtain HD contributions to
the effective field theory action for two M2-branes. We obtain 
several HD terms for two M2-branes, and we  also compare them  with an earlier study that has
been done using the component fields.

The remaining paper is organized as follows. In section $2$, we use derivative expansion of the effective
action of a real superfield in three dimensions to show that supersymmetric HD terms will not produce a
kinetic energy term for the auxiliary field. In section $3$, we  apply this formalism to calculate    
the HD terms for a  non-interacting supersymmetric  scalar field theory.
In section $4$, we use derivative expansion to generate HD terms with mass dimensions 
four, five and six for an interacting  supersymmetric theory. 
In section $5$, we are going to review the construction of the BLG theory in $\mathcal{N} = 1$
superspace formalism. In section $6$, we are going to analyse the  
effective field theory action for two  M2-branes. We will   generate
all  six dimensional topological HD terms for 
two M2-branes.  In the last section,  we will summarize our results and discuss few extensions of this 
work.

\section{Auxiliary Fields }
In this section, we use derivative expansion of the effective action of a real superfield in three
dimensions to show that supersymmetric HD terms will not produce a kinetic energy term for the
auxiliary field. So,   the theory, even after the HD terms have been added to it, 
can still be expressed in terms of physical fields. We will apply these results to a 
 non-interacting  real scalar superfield theory. 

Now before we  present our argument for the supersymmetric theories,  let us analyze  the leading order
  HD terms for a non-supersymmetric theory. 
  These terms will  correct the kinetic energy term in the low energy effective action   of the theory.
A natural framework to study a set of HD terms in a particular theory
is the derivative expansion of the low energy effective action,
which reproduces the theory in the infrared limit. We will now perform such an expansion for
a free massless scalar field  $(\phi)$ theory  in three dimensions.
It is important to list the mass dimension of various fields, since in effective field theories,
we consider the action up to a particular dimension,  which are suppressed by some microscopic length scale $l$.
Now from the kinetic terms
of various fields,  one can obtain the mass dimension for each field and derivative, 
$ [\phi]=1/2,\,\,[\partial ]=1,\,\,[m]=1.$ 
It is important to
state here that even if we are only interested in studying six dimensional  terms, 
we have to include four and five dimensional terms for the consistency of the low energy 
effective action expansion.

Now consider a derivative expansion of a low-energy effective action of a real 
superfield $\Phi$ in a
generic $\mathcal{N}=1$ supersymmetric theory 
\footnote{Here we use the notation of Ref.\cite{1001}}
\be S_t=\int d^2\theta d^3x {\cal L}(\Phi,D\Phi,...).\ee
Since the length scale $l$ controls the derivative expansion, 
the leading order term has three mass dimensions. After integrating over the fermionic coordinates we get
\be S_t=S_0+l\,S_1+l^2\,S_2+ O(l^3)\ee
where $S_n=S_n(\phi,\psi,F);\,\,\, n=0,1,2,...$ are functions of the component fields and 
have $n+3$ mass dimensions.
Therefore, $S_0$ must be at most quadratic in the auxiliary field $F$,
since $\Phi=\phi+\theta\psi -\theta^2F$, i.e., $[F]=3/2$. This means the general form of $S_0$ is given by
\be S_0=\int d^3x\, \left[\frac{\alpha}{2} F^2+g(\phi,\psi)\,F+k(\phi,\psi)\right].\ee
Now let us obtain the field equation of $F$, it reads
\be S_t'=S_0'+l\,S_1'+l^2\,S_2'+ O(l^3)=0\ee
where $S'={\delta S \over \delta F}$.
Since the microscopic scale $l$ controls the derivation expansion of the effective action
and suppresses HD terms, it is natural to expand the fields in terms of $l$. 
Also all field fluctuations larger than $1/l$ has been integrated out,
therefore, the only field fluctuations, we have to consider should be less than $1/l$.
It is natural to expand $F$ in terms of $l$
\be F=F_0+lF_1+l^2F_2+ O(l^3)\label{expandF}. \ee
Using equation (\ref{expandF}) and expand $S_1'$ and $S_2'$ in terms of $l$,  we get the following equations
\bea \alpha F_0+ g(\phi,\psi)=0,\nonumber \\
\alpha F_1+S_1'(\phi,\psi,F_0)=0,\nonumber \\
\alpha F_2+S_2'(\phi,\psi,F_0)+ S_1''(\phi,\psi,F_0) F_1=0.
\eea
 These equations show  that all the fields $F_n$'s, can be all expressed as functions of the
 physical fields $\phi$ and $\psi$.  Therefore, $F $ has no kinetic term,  and the action  can be written entirely
 in terms of $\phi$ and $\psi$. This result holds 
 independent of the form of the HD terms in $S_n$. In fact,
 one can extend this argument to the four dimensional chiral and vector supersymmetric  field theories in 
 $\mathcal{N}=1$ superspace  \cite{4dim}.

\section{Application}
Now we will apply the above argument to the supersymmetric action of a free massless real superfield 
$\Phi$ with higher derivative terms.
  Consider a real superfield   $\Phi (x, \theta) = \phi+ \theta^{a}\psi_\a-\theta^2 F $ \cite{1001},
whose action is given by
\be S_0= -  \int d^3x d^2\theta \left[{1 \over 2}\,(D^\a \Phi)^2\right]. \ee
In component form, the above action can be written as
\be S_0=\int d^3x {1 \over 2} [F^2+i\,\psi^\a\, {\pa_{\a}}^\b\, \psi_\b +\phi \Box \phi] .\ee
Using field equation of the auxiliary field $F =0$, which follows from this action, we can write  the action
  of the component fields as,
\be S_0=\int d^3x {1 \over 2}[i\,\psi^\a\, {\pa_{\a}}^\b\, \psi_\b +\phi \Box \phi] .\ee
Now considering the above action with four and five dimensions HD terms, we obtain the following total action,
\be  S_t=   \int d^3x\,\,d^2\theta \left[ \Phi D^2\Phi +\alpha\,l\, (D^2 \Phi)^2 +\beta\, l^2\, D^2\Phi \Box \Phi \right]+O(l^3),
\ee
where $\alpha$ and $\beta$ are some couplings. Expanding the above action in component one obtains 
\bea S_t &=& \int d^3x \,\,{1 \over 2} [F^2 + \phi \Box \phi +i \psi^\a {\partial_\a}^\b \psi_\b] +\alpha \,l\, [\psi^\b \Box \psi_\b +2F\,\Box \phi]
\nonumber\\
 && + \beta \, l^2\, [\phi \Box^2 \phi -i\,\psi_\a {\partial^\a}_\b \, \Box \psi^\b+F \Box F]+O(l^3).  \eea
It may be noted that the auxiliary field has a kinetic  term. Such terms occur 
in most supersymmetric field theories, if HD terms are considered. 
The field equation of the auxiliary field is given by
\be F+2\,\alpha\, l \Box \phi+\beta\,l^2\, \Box F=0. \ee
It is natural to expand $F$ in terms of $l$
\be F=F_0+lF_1+l^2F_2+ O(l^3). \ee
Using  field equation,  one gets,
\bea F_0&=&0, \nonumber\\  F_1&=&-2\alpha \Box\phi, \nonumber\\ F_2&=&-\beta\,\Box F_0=0. \eea
Now the total action reads
\bea S_t &=& \int d^3x \,\,  {1 \over 2} [\phi \Box \phi +i \psi^\a {\partial_\a}^\b \psi_\b] +\alpha \,l\, \psi^\b \Box \psi_\b   \nonumber\\
 &&+ l^2\, [(\beta-2\alpha^2)\phi \Box^2 \phi -i\,\beta\, \psi_\a {\partial^\a}_\b \, \Box \psi^\b]+O(l^3).  \eea

 It is not surprising that supersymmetry requires the existence of additional four dimension terms, and such
 terms did not exist in the non-supersymmetric version of the action. This is because the 
 field content of the supersymmetric theory   contains the fermionic field $\psi$.
 However, the interesting result here is that four dimension terms include an interesting fermionic topological term,
$ i\, \epsilon^{\mu\nu\sigma} \partial_{\mu} \psi_\b \partial_{\nu} \psi^{\g} \, ({\gamma_{\sigma}})^{\b}_{\g}.$
Even though, it is a total derivative for four mass-dimension
terms,  it can be argued   that 
it is possible to have   a five dimensional topological term in an
interacting theory. Furthermore, in the coming section, we show the possibility of having
a pure fermionic topological term with six mass dimensions of the form,  
$ \, \psi^2 \,\epsilon^{\mu\nu\sigma} \partial_{\mu} \psi_\b
\partial_{\nu} \psi^{\g} \, ({\gamma_{\sigma}})^{\b}_{\g}.$

\section{Interacting  Theory}

Now we will analyze  higher derivative terms for the  interacting supersymmetric $\phi^6$ theory, which is renormalizable in three dimensions.
The supersymmetric action for this theory can be written as
\be S_{0}= \int d^3x d^2\theta \left[-{1 \over 2} \,(D^\a \Phi)^2+{\lambda \over 4!} \Phi^4 \right].\ee
In component form, the above action can be written as
\bea S_{0}&=& \int d^3x  \,\,  {1 \over 2} [F^2+i\,\psi^\a\, {\pa_{\a}}^\b\, \psi_\b +\phi \Box \phi]
  \nonumber \\ &&   +{\lambda \over 2}
\phi^2\psi^2+  \,  {\lambda \over 3!}F \phi^3   . \eea
Using the field equation for the auxiliary field,   $ F=- \lambda   \phi^3/3!  $,
we obtain
\be S_{0}=\int d^3x \, \,  {1 \over 2} [i\,\psi^\a\, {\pa_{\a}}^\b\, \psi_\b +\phi \Box \phi] +{\lambda \over 2} \phi^2\psi^2-{\lambda^2 \over 2(3!)^2} \phi^6 .\ee
Now we will write the low energy effective theory with all possible HD terms consistent with the symmetries up
to dimension five. The list
of independent four and five dimensional terms is given by
\bea S_1&=&  l\, \int d^3x d^2\theta \,[ \, c_1 (D^2\Phi)^2+c_2 \Phi^2(D_{\alpha}\Phi)^2+c_4\Phi^6],
\nonumber\\
S_2&=&l^2\, \int d^3x d^2\theta \,[ \, c_5 D^2\Phi\Box\Phi +c_6\,\Phi^2(D^2\Phi)^2+c_7 \Phi D^2\Phi(D_\a \Phi)^2\nonumber\\
&& +c_8\, (D_{\alpha}\Phi)^4+c_9 \Phi^4(D_{\alpha} \Phi)^2+c_{10}\Phi^8].\eea

Now we can write all the four dimensional terms in component fields as
\bea \int d^2\theta\,(D^2\Phi)^2 &=&  \psi_\a  \Box {\psi}^{\a}+ 2 F\, \Box \phi,  \nonumber \\
  \int d^2\theta\,\Phi^2 (D_{\a}\Phi)^2 &=& \phi^2 \, [F^2+i\,\psi^\a\, {\pa_{\a}}^{\b}\, \psi_\b +\phi \Box \phi] +\psi^4+2 \psi^2\phi F, \nonumber \\
  \int d^2\theta\,\Phi^6&=& 30 \psi^2 \phi^4+6 F\phi^5. \eea
We can also write  all the five dimensional terms in component fields as
\bea \int d^2\theta\,(D^2\Phi)\, \Box \Phi &=&  \phi \Box^2 \phi -i\,\psi_\a {\partial^\a}_\b \, \Box \psi^\b+F \Box F,  \nonumber \\
 \int d^2\theta\,(D^2\Phi)^2\Phi^2 &= &\phi^2\,[\psi^\b \Box \psi_\b +i\, \epsilon^{\mu\nu\sigma} \partial_{\mu} \psi_\b \partial_{\nu} \psi^{\g}
\, ({\gamma_{\sigma}})^{\b}_{\g} \nonumber \\ && +2F\,\Box \phi] + 2\phi\,F^3 \nonumber \\ && +2\psi^2\, F^2+i8 \phi \psi_{\a} {\pa}^{\a}_{\b} \psi^{\b}F, \nonumber
\\
  \int d^2\theta\,(D^2\Phi)\Phi (D_{\a}\Phi)^2 & =  & F^2 \psi^2-i\psi^2\,\psi^\a {\pa}_{\a}^{\b}\psi_{\b}+\phi \Box \phi \,\psi^2
\nonumber \\ && +i \phi F\,\psi^{\a} {\pa}_{\a}^{\b} \psi_{\b}  -\phi \, \psi^{\g} {\pa}_{\mu}\psi_\g {\pa}^{\mu}\phi \nonumber \\ && +i\epsilon^{\mu\nu\sigma}
\phi \,\psi^{\b} ({\gamma_{\sigma}})^{\a}_{\b}{\pa}_{\mu} \psi_{\a}\pa_{\nu} \phi,
 \nonumber \\  \int d^2\theta\,(D_\a\Phi)^4 &=&   \psi^2 [\pa_{\mu}\phi \pa^{\mu}\phi+ 2 F^2],
 \nonumber \\
  \int d^2\theta\,\Phi^4 (D_{\a}\Phi)^2 &=  & 12\, \psi^4 \phi^2+ 4\,\psi^2\phi^3 F+8\, \psi^2 \phi^3F \nonumber\\
&& +\phi^4 [F^2+i\,\psi^\a\, {\pa_{\a}}^{\b}\, \psi_\b +\phi \Box \phi],\nonumber \\
  \int d^2\theta\,\Phi^8 &=&  56\,\psi^2 \phi^6+8\, \phi^7 F.  \eea

The term  $ \, \psi_\b ({\gamma_{\sigma}})^{\b}_{\g}\psi^\g\,\epsilon^{\mu\nu\sigma} 
 \partial_{\mu} \phi \partial_{\nu} \phi $      mixes the scalars and fermions in 
 five dimensional terms. Usually such topological terms are 
  are non-renormalizable. This   
   makes them an important tool to study
 the non-perturbative nature of the theory. 
 Although, this term vanishes identically for a single scalar superfield $\Phi$, 
 we will show in the next section that with more than one superfield (i.e., $\Phi^I$ is 
 in some representation of certain group $G$) these terms become  non-vanishing.

Now the list of six dimension terms in supersymmetric $\phi^6$ theory is given by
\bea S_3=&&l^3\, \int d^3x d^2\theta \,[ c_{11} (D^2 \Phi)^3 \Phi  + c_{12} (D^2\Phi)^2 \Phi^4+c_{13} (D^2\Phi) \Phi^7 \nonumber \\ && + c_{14} (D_{\alpha}\Phi)^2 \, D^2 \Phi \,\Phi^3 +  c_{15} (D_{\alpha}\Phi)^2 \, (D^2 \Phi)^2 \nonumber \\ &&  +c_{16}\,D^2 \Phi(\Box \Phi)\Phi^2    + c_{17}\,(\Box \Phi)^2 + c_{18}\,(\Box \Phi)\Phi^5   +c_{19}\, \Phi^{10}].\eea
We can   write  all the six dimensional terms in component fields as
 \bea \int d^2\theta\,(D^2\Phi)^3 \Phi =&&  3F^2\, [ i\psi^\a  \pa_{\a}^{\b} {\psi}_{\a}+  F^2+ \phi \Box \phi] \nonumber \\ && -6\, \phi F \,\pa_{\mu}\psi^{\a}\pa^{\mu}\psi_{\a} \nonumber \\ &&  +i6\, \phi F \,\epsilon^{\mu\nu\sigma} \pa_{\mu}\psi^{\a} (\gamma_{\sigma})_{\a \b}\pa_{\nu}\psi^{\b},  \nonumber\\
 \int d^2\theta\,(D^2\Phi)^2 \Phi^4 =&&  2F^3\, \Phi^3 + 3 \psi^2 \phi^2 F^2+ i8 \psi_{\a}  \pa^{\a}_{\b} {\psi}^{\b} \phi^3 F \nonumber \\ && + 2\phi^4 F \Box \phi    -{1 \over 2} \phi^4 \pa_{\mu} \psi^{\b} \pa^ {\mu} \psi_{\b} \nonumber \\ &&   -{i \over 2}\, \phi^4 \epsilon^{\mu\nu\sigma} (\gamma_{\sigma})^{\b}_{\a}\pa_{\mu}\psi_{\b}\pa_{\nu}\psi^{\a} \nonumber \\ && +4i F\phi^3 \psi^{\a}\pa_{\a}^{\b}\psi_{\b}, \nonumber\\
   \int d^2\theta\,(D^2\Phi) \Phi^7 = &&\phi^7 \Box \phi  -7i \phi^6 \psi^{\a}\pa_{\a}^{\b}\psi_{\b}+ 7 F^2 \phi^6\nonumber \\ && +42 \psi^2 F\phi^5,  \nonumber\\
  \int d^2\theta\,(D_{\a}\Phi)^2 (D^2\Phi) \Phi^3 =&&  \phi^3 \Box \phi \psi^2 -{i \over 4} \phi^3 \epsilon^{\mu\nu\sigma} (\gamma_{\sigma})^{\b}_{\a}\pa_{\mu}\psi_{\b}\pa_{\nu}\psi^{\a}\nonumber \\ && + {1 \over 4} \phi^3 \psi_{\b} \pa_{\mu} \psi^{\b} \pa^ {\mu} \phi   +{i \over 4}\, \phi^3 F \psi^{\a}\pa_{\a}^{\b}\psi_{\b} \nonumber \\ &&  +3i \phi^2 \psi^2 \psi^{\a}\pa_{\a}^{\b}\psi_{\b}   -2 F^3 \phi^3  \nonumber \\ &&
   + F \phi^3\, \pa_{\mu}\phi \pa^{\mu}\phi +{3 \over 2} F^2 \psi^2 \phi^2+ 6 F \psi^4 \phi, \nonumber\\
    \int d^2\theta\,(D_{\a}\Phi)^2 (D^2\Phi)^2 =&&  F \psi^2 \Box \phi -{i \over 2} \psi^2 [\pa_{\mu} \psi_{\a}\pa^{\mu} \psi^{\a}\nonumber \\ && +i \epsilon^{\mu\nu\sigma} (\gamma_{\sigma})^{\b}_{\a}\pa_{\mu}\psi_{\b}\pa_{\nu}\psi^{\a}]\nonumber\\ &&- {3 \over 2} F^2 \psi^{\a} \pa_{\a}^{\b} \psi_{\b}+{3 \over 2} F [\pa_{\mu} \psi_{\a} \pa^{\mu} \phi \psi^{\a}\nonumber \\  &&  +i \epsilon^{\mu\nu\sigma} (\gamma_{\sigma})^{\b}_{\a}\pa_{\mu}\psi_{\b}\pa_{\nu}\phi \psi^{\a}], \nonumber\\
  \int d^2\theta\,(D^2\Phi) \Box\Phi  \Phi^2 = &&  (\Box \phi)^2 \phi^2 +2i \psi_{\a}  \pa^{\a}_{\b} {\psi}^{\b} \phi \Box \phi \nonumber \\ && + F \Box F \phi^2 +2 \phi F  \psi^{\b} \Box \psi_{\b} \nonumber\\
   && +F \Box F \phi^2 +2 F^2  \phi \Box \phi -2 F  \psi^2 \, \Box \phi, \nonumber\\
   \int d^2\theta\,(\Box\Phi )^2  =  &&2 \Box F \Box \phi  + \Box \psi_{\a} \Box \psi^{\a}, \nonumber\\
    \int d^2\theta\,(\Phi )^{10}   =  &&{45 \over 2} \phi^8 \psi^2  + 10 F \phi^{9}, \nonumber\\
     \int d^2\theta\,(\Box\Phi ) \Phi^5  =  &&2 \phi^5 \Box F + 5 \phi^4 \psi_{\a} \Box \psi^{\a}
    \nonumber \\ && + 5 F \phi^4 \Box \phi + 20\, \phi^3 \psi^2 \Box \phi. \eea
    Notice the existence of non-vanishing purely fermionic topological terms as well as mixed
    topological terms. With more than one superfield, these terms should play an important role in 
    the effective action of M2-branes.

As we have mentioned before, having a single superfield forces certain topological terms to vanish as 
a result of anti-symmetrization of the  spacetime derivatives. Let us list few examples of those terms
keeping in mind that some of them (dimension 6 terms) will appear in the following section. Thus, we can write
the five dimensions term as
\bea &&\int d^2\theta\,(D_{\b}\Phi) (D^{\b}D_{\rho}\Phi) (D^{\rho}\Phi) \Phi \nonumber\\
&&\supset \epsilon^{\mu\nu\sigma}
{\pa}_{\mu} \phi {\pa}_{\nu} \phi \,\psi^{\b} ({\gamma_{\sigma}})^{\a}_{\b}\psi_{\a}+\epsilon^{\mu\nu\sigma}
\phi \,{\pa}_{\mu} \phi {\pa}_{\nu} \phi {\pa}_{\sigma} \phi,  \eea
 and the six dimensional term as
\bea &&\int d^2\theta\, \Phi^3 (D_{\b}\Phi) (D^{\b}D_{\rho}\Phi) (D^{\rho}\Phi)  \nonumber\\
&& \supset \epsilon^{\mu\nu\sigma}
\phi^2{\pa}_{\mu} \phi {\pa}_{\nu} \phi \,\psi^{\b} ({\gamma_{\sigma}})^{\a}_{\b}\psi_{\a}+\epsilon^{\mu\nu\sigma}
\phi^3 \,{\pa}_{\mu} \phi {\pa}_{\nu} \phi {\pa}_{\sigma} \phi. \eea

\section{Action for M2-Branes}
In this section,  we are going to review the construction of the action 
for two M2-branes. The BLG theory describes the physics of two M2-branes.
So,  here  we  review the construction of the BLG theory in 
$\mathcal{N} = 1$ superspace formalism   \cite{n1}. The 
gauge  fields in the BLG theory are  valued in a Lie $3$-algebra 
rather than a conventional Lie algebra. A Lie $3$-algebra is
a vector space endowed with a trilinear product,
\begin{equation}
[T^a,T^b,T^c] = f^{abc}_d T^d. 
\end{equation}
The  structure constants of this Lie $3$-algebra are totally antisymmetric
  in $a,b,c$. They also  satisfy the Jacobi identity  \cite{blgblg},
\begin{equation}
f^{[abc}_g f^{d]eg}_h = 0.
\end{equation}
The metric of this Lie $3$-algebra can be defined by  taking  the trace over the Lie $3$-algebra indices,
\be
h^{ab} = Tr(T^a T^b).
\ee
It is also possible to define a symmetrised trace of four Lie $3$-algebra generators as
\be
Str (T^a T^b T^c T^d)= m \, h^{(ab}h^{cd)},
\ee
where $m$ is a constant. For the Lorentz Lie $3$-algebra, it
  is possible to consider a set of generators corresponding to a compact subgroup of the full symmetry group. Hence, we can choose  the generators of   a $SU(2)$ Lie algebra,   and write  \cite{mukhi-1}
\begin{eqnarray} Tr(T^a T^b)&=&{1\over 2}\delta^{a b}, \nonumber \\
  STr(T^a T^b T^c T^d)&=&{1\over 4}\delta^{(a b}\,\delta^{c d)}.   \end{eqnarray}

The gauge fields are valued in the Lie $3$-algebra, $\Gamma^{\alpha}_{\phantom{\alpha}ab} T^a T^b = \Gamma^{\alpha} $.
The BLG theory has been written using  ${\cal N}=1$ superspace formalism. 
This is done by writing defining 
${\Phi^{I}}_a={\phi^I}_a+\theta^{\a}\psi^I_{\a a}-\theta^2{F^{I}}_a,$ with $  I=1,2,..8,$
where $a$ is the three-algebra index
with a structure constant $f^{a b c d}$.     So, we can write the action for the BLG theory as \cite{n1}
 \begin{eqnarray}
S_0 & = &-  \int d^3 x \, d^2 \theta \,  \Big[\, \frac{1}{4} \left(
D^{\alpha}\Phi^{I}_{d}    + \, f
^{abc}_{\phantom{abc}d}\,\Gamma^{\alpha}_{\phantom{\alpha}ab}
\Phi^{I}_{c} \right)^2 \nonumber \\ &&  +\, \,\frac{1}{8} \,f^{abcd}\,(D^{\alpha}
\Gamma^{\beta}_{\phantom{\beta}ab})(D_{\beta} \Gamma_{\alpha \,cd})
  \,\nonumber\\
\label{action}
 & & + \frac{1}{6} \,f^{cda}_{\phantom{cda}g} f^{efgb}\, (D^{\alpha}
\Gamma^{\beta}_{\phantom{\beta}ab})\Gamma_{\alpha \,c d}\Gamma_{\beta
\, e f}   \nonumber \\  &&   +\,  \frac{1}{24} \,f^{abcd} \,C_{IJKL}
\,\Phi^{I}_{a}\Phi^{J}_{b}\Phi^{K}_{c}\Phi^{L}_{d}
\,\Big].
\end{eqnarray}
The component field definitions are as
\begin{eqnarray}
\Phi^I_a | = \phi^I_a, && D_{\alpha}\Phi^I_a | = \psi^I_{\alpha \,a},
 \nonumber   \\  D^2 \Phi^I_a | -
=F^I_a, &&   \Gamma_{\alpha \,ab} |=
\chi_{\alpha\,ab},    \nonumber   \\ \frac{1}{2}\, D^{\alpha}
\Gamma_{\alpha\,ab} |=  B_{ab}, &&
\Gamma_{\alpha\,ab} | =  2 \lambda_{\alpha\,ab} -
i\,\partial^{\,\beta}_{\alpha}\chi_{\beta\,ab},
  \nonumber   \\ D_{\alpha}\Gamma^{\beta}_{\,\,a b}| =
i \left(\gamma_{\mu} \right)^{\,\beta}_{\alpha}
A^{\mu}_{\,\,a b} - \delta^{\,\beta}_{\alpha} B_{a b}, &&  D^2
\Gamma^{\alpha}_{\,\,a b} | = 2 \lambda_{\,\,a b}^{\alpha} +
i\,\partial_{\,\,\beta}^{\alpha}\chi_{\,\,a b}^{\beta},  \nonumber \\
D^{\alpha}\Gamma_{\beta \,a b}|=
i \left(\gamma_{\mu} \right)_{\,\beta}^{\alpha}
A^{\mu}_{\,\,a b} + \delta_{\,\beta}^{\alpha} B_{a b}, &&
\frac{1}{2} D^{\beta}D_{\alpha} \Gamma_{\beta \, ab}| =
\lambda_{\alpha \, ab}.
\end{eqnarray}
An octonion algebra
 $ \{1,\, e_i \},  $ with $i=1,..,7$,
such that
 $
e_i e_j=c_{ijk}\,e_k - \delta_{ij},
 $ has been used to defined $C_{IJKL}$. This is done by taking a
 totally antisymmetric tensor $c_{ijk}$.
The seven dimensional dual of this  is
 $
c_{ijkl} = \frac{1}{6} \,\epsilon_{ijklmno}\,c^{mno}
 $.  Now it is possible to construct an $SO(7)$ invariant
tensor $C_{IJKL}\,\,_{I,J,K,L=1,...,8}$ which is self dual in eight
dimensions,
$
C_{ijk8}=c_{ijk}, C_{ijkl}=c_{ijkl}
$. This   octonionic structure constants can be used to construct $SO(8)$
 gamma matrices \cite{n1}. So, we can use  write
$
 (\Gamma^i)_{A\dot{A}} = c^i_{\,\,A\dot{A}} +
\delta_{8\dot{A}}\delta_{Ai} - \delta_{8A} \delta_{\dot{A}i}$,  where $
  i=1,...,7 $ and $  A,\dot{A}=1,...,8$.  We also have $ (\Gamma^8)_{A\dot{A}}
= \delta_{A\dot{A}}
c^{i}_{8\dot{A}} = c^{i}_{A8}= 0.
$
Here we have defined $\hat{\Gamma}^I_{\dot{A}A}=\left(\Gamma^{T}\right)^I_{\dot{A}A}$,  and
$\Gamma^I\hat{\Gamma}^J + \Gamma^J\hat{\Gamma}^I =
2\delta^{IJ}$. Now the clifford algebra can be written as
\be
\gamma^I\gamma^J+\gamma^J\gamma^I= 2 \delta^{IJ},
\ee
where
\be
\gamma^I=\left(\begin{array}{cc}0 & \Gamma^I_{\,\,A\dot{A}} \\
\hat{\Gamma}^I_{\,\,\dot{A}A} & 0\end{array}\right)
\,.
\ee
Now we can also write
\bea
\Gamma^{IJ}_{\,\,AB} &=&\frac{1}{2}\left(\Gamma^I_{A\dot{A}}\hat{\Gamma}^J_{\dot{A}B}-\Gamma^J_{A\dot{A}}\hat{\Gamma}^I_{\dot{A}B}\right)\nonumber \\ &=& C^{IJ}_{\,\,\,AB} +\delta^I_A \delta^J_B
-\delta^I_B \delta^J_A\,.
\eea

\section{Topological   Terms for M2-Branes}
In this section, we will analyse the effective action for two M2-branes. We will  
use the above analysis to argue for the existence of new topological
HD terms in the effective action of M2-branes. 
The effective M2-brane action can be expanded in terms of Planck length $l_p$ as follows
\be S_{BLG}=S_0+l_p^3 S_3+...\ee
Therefore, the first correction to the leading contribution is of dimension six. This is why we have expanded
our effective action in the general $\phi^6$ theory up to such  order.
  The action of the  theory without the gauge field, i.e., the Higgs branch, is given
by \be S_0=- \int dx^3 d^2\theta \,Tr\left( (D_{\a} \Phi)^2 +{1\over 12} [\Phi^I,\Phi^J,\Phi^K]\Phi^L C_{IJKL} \right).\ee
Since the leading correction is of order $O(l_p^3)$, the field equation for the  auxiliary field is given by
\be F^I_a={-1 \over 6} f^{a b c d} C^I_{JKL} \Phi^J_b \Phi^K_c\Phi^L_d + O(l_p^3). \label{aux}\ee

The non-vanishing six dimensional topological terms can be classified as  bosonic, fermionic and mixed
terms. Here we list all such terms.
We find the following bosonic terms,
\bea {\cal L}_{b1}&= &\int d^2\theta \, STr\left( C_{IJKL}C^I_{J'K'L'} \, \Phi^{J'K'L'}D_{\b} \Phi ^J D^{\b}D_{\g}\Phi^K D^{\g}\Phi^L \right)\nonumber\\
&& \supset STr\left( C_{IJKL}C^I_{J'K'L'} \, \phi^{J'K'L'}\epsilon^{\m \n \sigma }\pa_{\m}\phi ^J \pa_{\sigma}\phi^K \pa_{\n}\phi^L \right)\nonumber\\
&=& {1\over 4} C_{IJKL}C^I_{J'K'L'} \, \phi^{J'K'L'}_{a' b' c'} {f^{a' b' c'}}_d\,   \epsilon^{\m \n \sigma } \nonumber \\ && \times  \pa_{\m}\phi_a^J \pa_{\sigma}\phi_b^K
\pa_{\n}\phi_c^L \, \delta^{(ab} \delta^{cd)} \eea
Another term that produces the same topological term is given by
\bea {\cal L}_{b1'}&= & \int d^2\theta \, STr\left( C_{IJKL}\, D^2\Phi^I D_{\b} \Phi ^J D^{\b}D_{\g}\Phi^K D^{\g}\Phi^L \right) \nonumber\\
&& \supset STr \left( C_{IJKL}F^I \, \epsilon^{\m \n \sigma }\pa_{\m}\phi ^J \pa_{\sigma}\phi^K \pa_{\n}\phi^L \right).\eea The relation between
this term and the previous one is clear upon using Eq.  (\ref{aux}). 
This term has also been obtained in earlier works on M2-branes \cite{mukhi-1}.
We also find  three different fermionic terms
\bea {\cal L}_{f1}&= &\int d^2\theta \, Tr \, \left( C^{IJKL}D^{\a}\Phi^I [D_{\a} \Phi ^J, D^{2}\Phi^K, D^{2}\Phi^L]\, \right)\nonumber\\
&& \supset  C^{IJKL}f^{abcd}\,\epsilon^{\m \n \sigma }\,\psi^I_a\cdot \psi^J_b \, \pa_{\m}\psi^K_c \cdot \gamma_{\n}\cdot\pa_{\sigma}\psi^L_d
\nonumber \\ {\cal L}_{f2}&= &\int d^2\theta \, STr\left( D^{\a}\Phi^{I} D_{\a} \Phi ^{J} D^{2}\Phi^{I} D^{2}\Phi^{J}\, \right) \nonumber\\
&& \supset  \delta^{(ab}\delta^{cd)}\,\epsilon^{\m \n \sigma }\,\psi^I_a\cdot \psi^J_b \, \pa_{\m}\psi^I_c \cdot \gamma_{\n}\cdot\pa_{\sigma}\psi^J_d
\nonumber \\ {\cal L}_{f3}&= &\int d^2\theta \, STr \, \left( D^{\a}\Phi^{I} D_{\a} \Phi ^{I} D^{2}\Phi^{J} D^{2}\Phi^{J}\, \right)\nonumber\\
&& \supset  \delta^{(ab}\delta^{cd)}\,\epsilon^{\m \n \sigma }\,\psi^I_a\cdot \psi^I_b \, \pa_{\m}\psi^J_c \cdot \gamma_{\n}\cdot\pa_{\sigma}\psi^J_d, \eea
where $\psi\cdot \xi={1\over 2} \psi^\a \xi_\a$ and  $\psi\cdot \gamma^\mu \cdot \xi= \psi_\a (\gamma^{\mu})^{\a \b }\xi_\b$.
This last term can be expressed in the notation of \cite{mukhi-1} as
$Tr \left( \bar{\Psi} \Psi \epsilon^{\m \n \sigma }D_{\m}\bar{\Psi}\gamma_{\n}D_{\sigma}\Psi\right).$
Finally, we find the following mixed terms,
\bea {\cal L}_{m1}&= &\int d^2\theta \, Tr \, \left( D_{\g}D_{\a}\Phi^J [\Phi ^I, D_\b \Phi^J, D^\b D^\g D^\a \Phi^I]\, \right)\nonumber\\
&& \supset  f^{abcd}\,\epsilon^{\m \n \sigma }\,\phi^I_a \phi^J_b \, \pa_{\m}\psi^I_c \cdot \gamma_{\n}\cdot\pa_{\sigma}\psi^J_d
\nonumber \\   {\cal L}_{m2}&= &\int d^2\theta \, STr \, \left(  C_{IJKL}\, D_{\g}D_{\a}\Phi^I \, \Phi ^J\, D_\b \Phi^K \,D^\b D^\g D^\a \Phi^L \, \right) \nonumber\\
&& \supset  C_{IJKL}\,\delta^{(ab}\delta^{cd)}\,\epsilon^{\m \n \sigma }\,\phi^I_a \phi^J_b \, \pa_{\m}\psi^K_c \cdot \gamma_{\n}\cdot\pa_{\sigma}\psi^L_d
\nonumber \\  {\cal L}_{m3}&= &\int d^2\theta \, Tr \, \left( {C^{IJK}}_{L}{C_{I'J'K'}}^L\, \Phi^{I'J'K'} [\Phi^I, \, D^\a\Phi ^J, D^2D_\a \Phi^K] \right) \nonumber\\
&& \supset  {C^{IJK}}_{L}{C_{I'J'K'}}^L\, \phi^{I'J'K'}_{a'b'c'}f^{a'b'c'd}{f^{abc}}_d\,\epsilon^{\m \n \sigma }\, \nonumber \\ && \times \phi^I_a \, \pa_{\m}\psi^J_c \cdot \gamma_{\n}\cdot\pa_{\sigma}\psi^K_d \eea
The same topological term can be produced by
\bea {\cal L}_{m3'}&= &\int d^2\theta \, Tr \, \left( {C^{IJK}}_{L}D^2\Phi^L \, [\Phi^I , D^\a\Phi ^J, D^2D_\a \Phi^K] \right) \nonumber\\
&& \supset  {C^{IJK}}_{L} \,F^L_d\,{f^{abc}}_d\,\epsilon^{\m \n \sigma }\,   \phi^I_a \, \pa_{\m}\psi^J_c \cdot \gamma_{\n}\cdot\pa_{\sigma}\psi^K_d \eea
The last mixed term can be written as
\bea {\cal L}_{m4}&= &\int d^2\theta \, STr \, \left( {C_{I'J'K'}}^I\, \Phi^{I'J'K'} \Phi^J \, D^\a\Phi ^I\, D^2D_\a \Phi^J \right) \nonumber\\
&& \supset  {C_{I'J'K'}}^I\, \phi^{I'J'K'}_{a'b'c'}{f^{a'b'c'}}_{d}\delta^{(ab}\delta^{cd)}\,\epsilon^{\m \n \sigma }\,\nonumber \\ && \times \phi^J_a \, \pa_{\m}\psi^I_b \cdot \gamma_{\n}\cdot\pa_{\sigma}\psi^J_c \eea
Another term that produces the same topological term, after using Eq. (\ref{aux}), takes the form
\bea {\cal L}_{m4'}&= &\int d^2\theta \, STr\, \left( D^2\Phi^I \Phi^J \, D^\a\Phi ^I\, D^2D_\a \Phi^J \right) \nonumber\\
&& \supset  \delta^{(ab}\delta^{cd)}\,F^I_d\,\epsilon^{\m \n \sigma }\,\phi^J_a \, \pa_{\m}\psi^I_b \cdot \gamma_{\n}\cdot\pa_{\sigma}\psi^J_c. \eea

It is worth mentioning here that these six dimensional HD terms of the M2-branes effective action have 
been calculated in earlier studies
  \cite{mukhi-1}. This was done by using a novel Higgs mechanism, and this reduced the M2-brane action
  to a matter-Yang-Mills theory describing the low energy effective action of multiple D2-branes.
The HD terms obtained in this work are written in terms of component fields, and the lowest order field 
equations in derivative expansion was used. This made several of these HD terms to vanish.
In contrast to this, the HD terms in this work have been constructed using superspace formalism, and  
they have been written in terms of $\mathcal{N}=1$ superfields.
This explains why all these terms were not obtained in earlier studies \cite{mukhi-1}.
In fact, in the earlier component formalism only the  pure bosonic topological term was obtained. 
Apart from this pure
bosonic terms, all the HD terms produced here using the superspace formalism, vanished upon using
lowest order field equations  in derivative expansion. The reason is that these terms contain fermions 
and can be written in terms of one or more factors
of $i\pa_{\a \b}\psi^{\b}$, which are set to zero by the lowest order field equation in derivative expansion. 
This explains why these terms were absent in  earlier studies  \cite{mukhi-1}.
The importance of these HD terms come from the fact that they modify the interaction Lagrangian, therefore,
they affect loop calculations of scattering amplitudes in the low energy effective theory \cite{qa, q1}. 
In addition, if
we consider path integral quantization for such an effective theory we have to sum over all such HD terms in our 
Lagrangian, otherwise we will not have the correct interaction Lagrangian \cite{pa, ap}.
\section{Conclusion}
In this paper, we have analyzed  the higher derivative terms for three dimensional supersymmetric theories with
$\mathcal{N} =1$ supersymmetry. We first analyzed the higher derivative terms for a general scalar superfield 
theory and demonstrated  that the
  auxiliary field will not acquire a kinetic term for all possible actions of the theory. Therefore, the theory 
  is completely describable in
  terms of its original field content. We calculated all four, five and six dimensional terms for such a theory 
  demonstrating the existence of
  various interesting topological terms.
  We obtained pure bosonic, pure fermionic and 
  terms which mix bosonic  and fermionic fields. 
  We also analyse the  effective action for the 
  BLG theory   in $\mathcal{N}=1$ superspace formalism. 
  We show the existence of several
  mixed and pure fermionic terms which vanish  upon using the lowest order field equations in derivative expansion.
  These terms were absent in the list of six dimensional terms generated in earlier studies \cite{mukhi-1}. 
 It may be noted that even though these terms vanish upon using
 the lowest order field equations in derivative expansion, it is important to consider them, as they can 
  affect loop calculations of scattering amplitudes in the low energy effective theory. 

 It will be interesting to generalize the results of this paper, for theories with higher amount of supersymmetry.
 Furthermore, it will be interesting to perform a similar analysis for matter fields coupled to gauge fields.
 The results thus obtained
 can be used for analyzing HD corrections to the M2-branes effective actions using the ABJM theory.
It is possible to extend this work done on global supersymmetry with higher derivative terms to local supergravity 
theories.
In fact, the supergravity extension  of scalar field theories with higher derivative terms
 has  been studied \cite{sg}. This analysis was done using supergravity in $\mathcal{N} =1$ superspace
 formalism. The elimination of auxiliary fields modifies both the kinetic and potential terms in this theory. 
 In this case, it has been demonstrated that potential energy can be generated even if there was no original 
 superpotential term in the action. It will be interesting to extend the results of this paper to
 local supergravity theories in three dimensions.

\end{document}